# Electric field sensor based on electro-optic polymer refilled silicon slot photonic crystal waveguide coupled with bowtie antenna

Xingyu Zhang*[a], Amir Hosseini*[b], Xiaochuan Xu[a], Shiyi Wang[c], Qiwen Zhan[c], Yi Zou, Swapnajit Chakravarty, and Ray T. Chen*[a]

[a] The University of Texas at Austin, 10100 Burnet Rd, MER 160, Austin, TX 78758;

[b] Omega Optics, Inc., 10306 Sausalito Drive, Austin, TX 78759;

[c] University of Dayton, 300 College Park, Dayton, OH 45469.

## ABSTRACT

We present the design of a compact and highly sensitive electric field sensor based on a bowtie antenna-coupled slot photonic crystal waveguide (PCW). An electro-optic (EO) polymer with a large EO coefficient, $r_{33}$=100pm/V, is used to refill the PCW slot and air holes. Bowtie-shaped electrodes are used as both poling electrodes and as receiving antenna. The slow-light effect in the PCW is used to increase the effective in-device $r_{33}$>1000pm/V. The slot PCW is designed for low-dispersion slow light propagation, maximum poling efficiency as well as optical mode confinement inside the EO polymer. The antenna is designed for operation at 10GHz.

## 1. INTRODUCTION

Electric field measurements play a crucial role in various scientific and technical areas, including process control, electric field monitoring in medical apparatuses, ballistic control, electromagnetic compatibility measurements, microwave integrated circuit testing, and detection of directional energy weapon attack. Conventional electronic electric field measurement systems currently use large active metallic probes, which can disturb the measured electromagnetic field and make sensors sensitive to electromagnetic noise. Photonic electric field sensors, on the other hand, are compact in size and can achieve broad bandwidth operation with minimal disturbance to the field to be measured. In this paper, we propose an antenna-coupled photonic crystal waveguide (PCW) based electric field sensor.

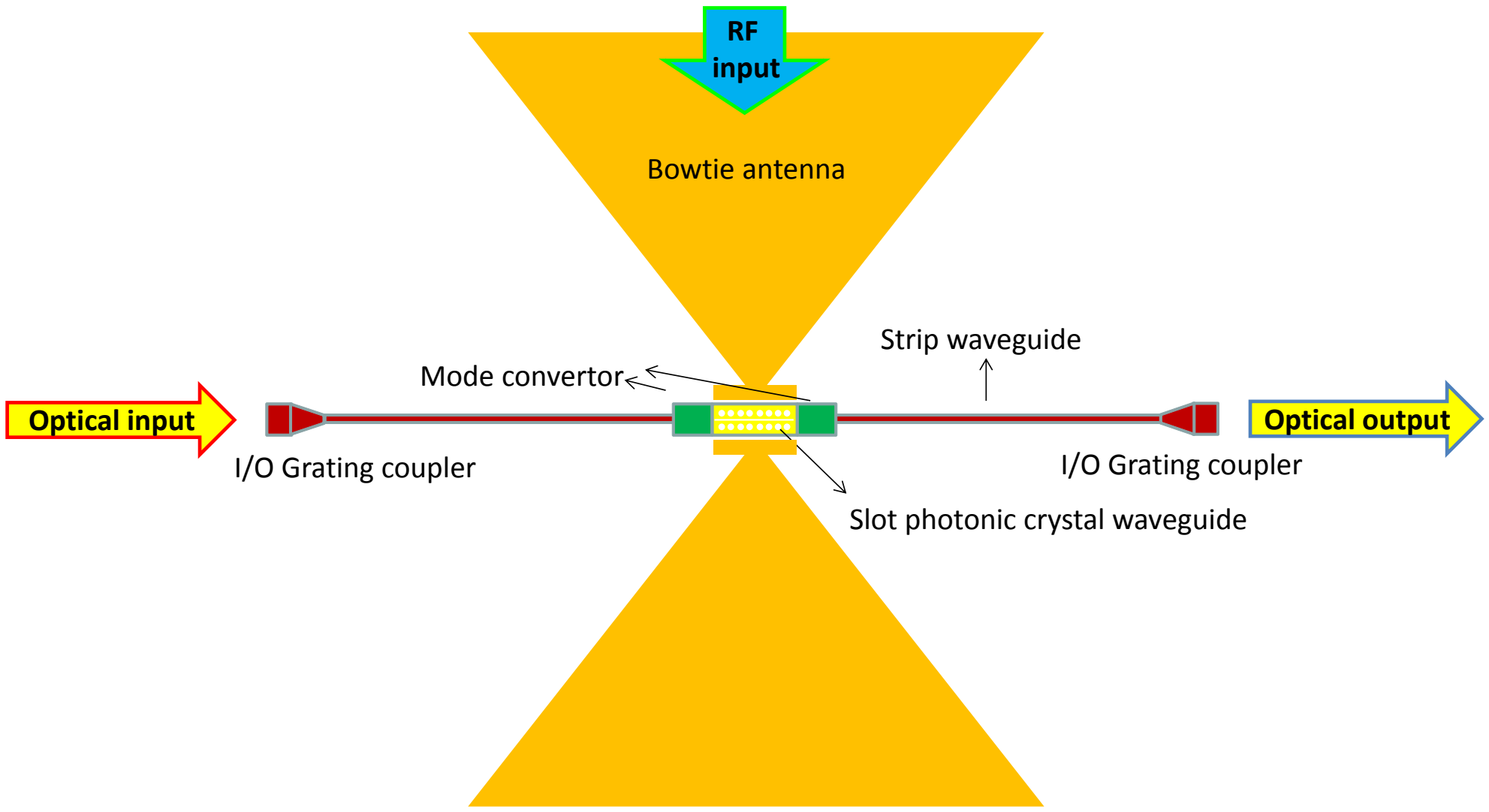

Figure 1. The schematic view of the photonic crystal modulator coupled with a bowtie antenna.

*xzhang@utexas.edu; phone 1 512-471-4349; fax 1 512 471-8575
*amir.hosseini@omegaoptics.com; phone: 1 512 996 8833; 1 512 873 7744
*raychen@uts.cc.utexas.edu; phone 1 512-471-7035; fax 1 512 471-8575

The slot PCW is band-engineered to achieve low-dispersion slow light propagation over a broad wavelength range and to increase the effective in-device electro-optic (EO) coefficient ($r_{33}$) >1000pm/V for low half wave switching voltage ($V_{\pi}$) ~1V. An EO polymer, SEO125 with a large $r_{33}$=125pm/V from Soluxra, LLC, is used to refill the PCW slot and air holes. The slot photonic crystal waveguide is designed for low-dispersion slow light propagation, maximum poling efficiency as well as optical mode confinement inside the EO polymer. A bowtie-shaped antenna is designed for operation at 10GHz and used as receiving antenna as well as poling electrode. Our electric field sensor consisting of a slot photonic crystal waveguide modulator coupled with a bowtie antenna is shown in Figure 1. In the next two sections, the design of slot photonic crystal waveguide modulator and the bowtie antenna will be discussed, respectively.

## 2. OPTICAL WAVEGUIDE DESIGN

### 2.1 Band engineered Electro-optic polymer refilled slot photonic crystal waveguide

Photonic crystal waveguides (PCWs) can be band-engineered to achieve low-dispersion slow light propagation in different ways [2]. In this work we have chosen the lateral lattice shifting approach [3] for the following advantages. First, all the PCW holes are the same and this increases the fabrication yield and reproducibility compared to techniques that require precise control of multiple hole diameters. Second, it facilitates targeting a desired group velocity over a bandwidth of interest since these two parameters can be tuned relatively independently compared to longitudinal lattice shifting. Third, it does not change the defect line width and facilitates efficient coupling between the fast-light mode of a silicon slot waveguide (group index ~3) and the slow-light mode in the PCW (group index>10).

A schematic of the band engineered slot PCW is shown in Figure 2 (a). The first three adjacent rows on each side of the defect line are shifted parallel to the line defect to modify the dispersion diagram of the defect mode. Using Rsoft BandSolve module we simulate the fundamental guided defect mode profile and band structure shown in Figure 2 (a)-(c). The PCW is designed to be fabricated on a silicon-on-insulator (SOI) substrate with 250nm top silicon and 3μm oxide layers. The PCW holes and slot are assumed to be filled with an EO polymer with a refractive index of 1.63. For lattice constant, a=425nm, it is found that with hole diameter, d=300nm, $s_1$=0, $s_2$=-85nm, $s_3$=85nm, slot width of $S_w$=320nm, and dW=1.54($\sqrt{3}$)a, we can achieve an average group index ($n_g$) of 20.4 ($\pm$10%) over 8.2nm bandwidth as shown in Figure 3 (d). Note that the second and third rows are shifted in different directions. The absolute value of the group velocity dispersion (GVD) remains below 10ps/nm/mm over the entire bandwidth, as shown in Figure 3 (e).

In order to efficiently couple light in and out of the device, a PCW taper has been designed (a=425nm, d=300nm, $s_1$=0, $s_2$=0, $s_3$=0, $S_w$=320nm, dW=1.45($\sqrt{3}$)a. Figure 3(c) and (d) show the band structures and the group index variations of the band engineered PCW and the PCW taper, respectively. A group index of ~6 of the PCW taper over the optical bandwidth of interest provides an effective interface for coupling between the slow light mode of the PCW (group index =20.4) and the fast light in the silicon slot waveguides (group index ~3).

The required length for an optical modulator to achieve π phase is given as $L=1/(2\sigma)\times(n/\Delta n)\times\lambda/n_g$= 255.9μm, where σ=0.33 is the fraction of the energy in the slot which is calculated over one complete period of the fundamental guided defect mode profile in the band engineered PCW in the BandSolve simulations in Figure 2(b), n=1.63 is the index of the electro-optics polymer, Δn=0.000733 is the change in the effective index of the EO polymer when voltage V=1V is applied, and λ=1.55μm is the free-space wavelength and $n_g$=20.4 is the group index. The change in the EO index is calculated using $\Delta n=n_3\ r_{33}\ V\ /\ (2S_w)$, where $r_{33}$=100pm/V is the EO coefficient of the polymer. Given the potentially large $r_{33}$=150pm/V of EO polymer SEO125 and demonstrated high poling efficiency achievable in wide slots (320nm compared to conventional 100nm) [4], the estimated $r_{33}$=100pm/V here is a realistic value. Therefore, from these calculations, the voltage required for π phase shift, $V_{\pi}$=1V, and thus $V_{\pi}\times L$= 1V×255.9μm=0.0256V-cm. Such a small $V_{\pi}\times L$ promises a compact and sensitive antenna coupled phase modulator. The length of the active section of the PCW is chosen to be 300μm in the design. The expected effective in-device $r_{33}$ is then calculated as

$$r_{33,effective} = \frac{\lambda S_w}{n^3 V_\pi \sigma L} = 1356\text{pm/V}$$

where, λ=1.56 μm, $S_w$=320nm (slot width), n=1.63 (polymer), L=225.9 μm, σ=0.33 is the confinement factor in the slot calculated by simulation.

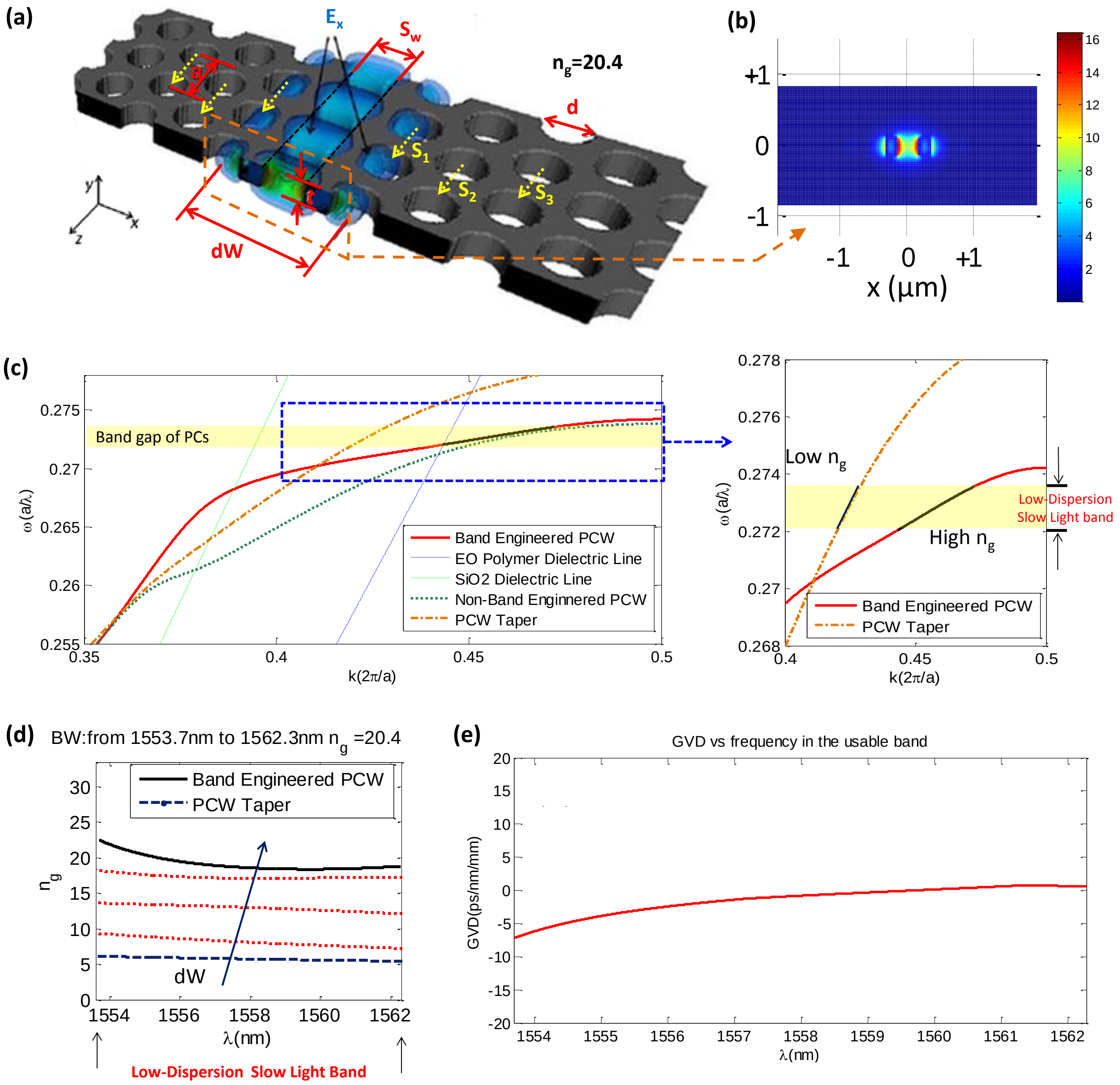

Figure 2. (a) A 3D schematic of band-engineered slot PCW, overlaid with the 3D electric field profile of the fundamental guided defect mode. (b) 2D view of the electric field profile of the fundamental guided defect mode in the band-engineered PCW at τ=0. See [1] for the definition of τ. (c) The band structure (normalized frequency v.s. normalized propagation constant, the fundamental guided defect mode) for 3 PCWs, the band-engineered PCW (a=425nm, d=300nm, $s_1$=0, $s_2$=-85nm, $s_3$=85nm, $S_w$=320nm, dW=1.54($\sqrt{3}$)a), the PCW taper (a=425nm, d=300nm, $s_1$=0, $s_2$=0, $s_3$=0, $S_w$=320nm, dW=1.45($\sqrt{3}$)a), and a non-band-engineered PCW (a=425nm, d=300nm, $s_1$=0, $s_2$=0, $s_3$=0, $S_w$=320nm, dW=1.54($\sqrt{3}$)a). The "flattening" of the mode in the band-engineered PCW compared to the non-band-engineered PCW can be noticed. The black curve highlights the low dispersion slow light section of the mode of the band-engineered PCW. The dielectric light lines corresponding to the $SiO_2$ (n=1.45) and EO polymer (n=1.63) cladding layers are shown. The useful part of the mode falls below the both light lines. (d) Variations of the group index vs. wavelength for the band-engineered PCW and the PCW taper. (e) Variations of the group velocity dispersion v.s. wavelength for the band-engineered PCW.

### 2.2 Photonic crystal waveguide coupler

In order to reduce the total insertion loss, an efficient coupling scheme between the silicon ridge waveguide and the PCW is necessary. Therefore, an efficient PCW coupler consisting of a mode converter and a PCW taper (group index taper) is designed [5], as shown in Figure 3 (a). The mode converter changes the mode profile of the input waveguide to that of the PCW as shown in Figure 3 (b) and (c). The PCW taper gradually increases (slows down) the group index (propagating light) from the interface with the mode converter to the interface with the high $n_g$ PCW. The PCW Taper consists of non-band engineered PCW (a=425nm, d=300nm, $s_1$=0, $s_2$=0, $s_3$=0, $S_w$=320nm), for which, the width of the line defect (dW) parabolically increases from dW=1.45($\sqrt{3}$)a to dW=1.54($\sqrt{3}$)a. The slot width ($S_w$), hole diameter (d), and the period (a) remain constant. In addition, sub-wavelength grating (SWG) couplers (in Figure 3(d)) are designed for an oxide bottom cladding and a SU-8 (polymer) top cladding, to couple light from input/output (IO) fibers into and out of the silicon ridge waveguide as shown in Figure 1. The maximum coupling efficiency of each grating is about 26% as shown in Figure (e).

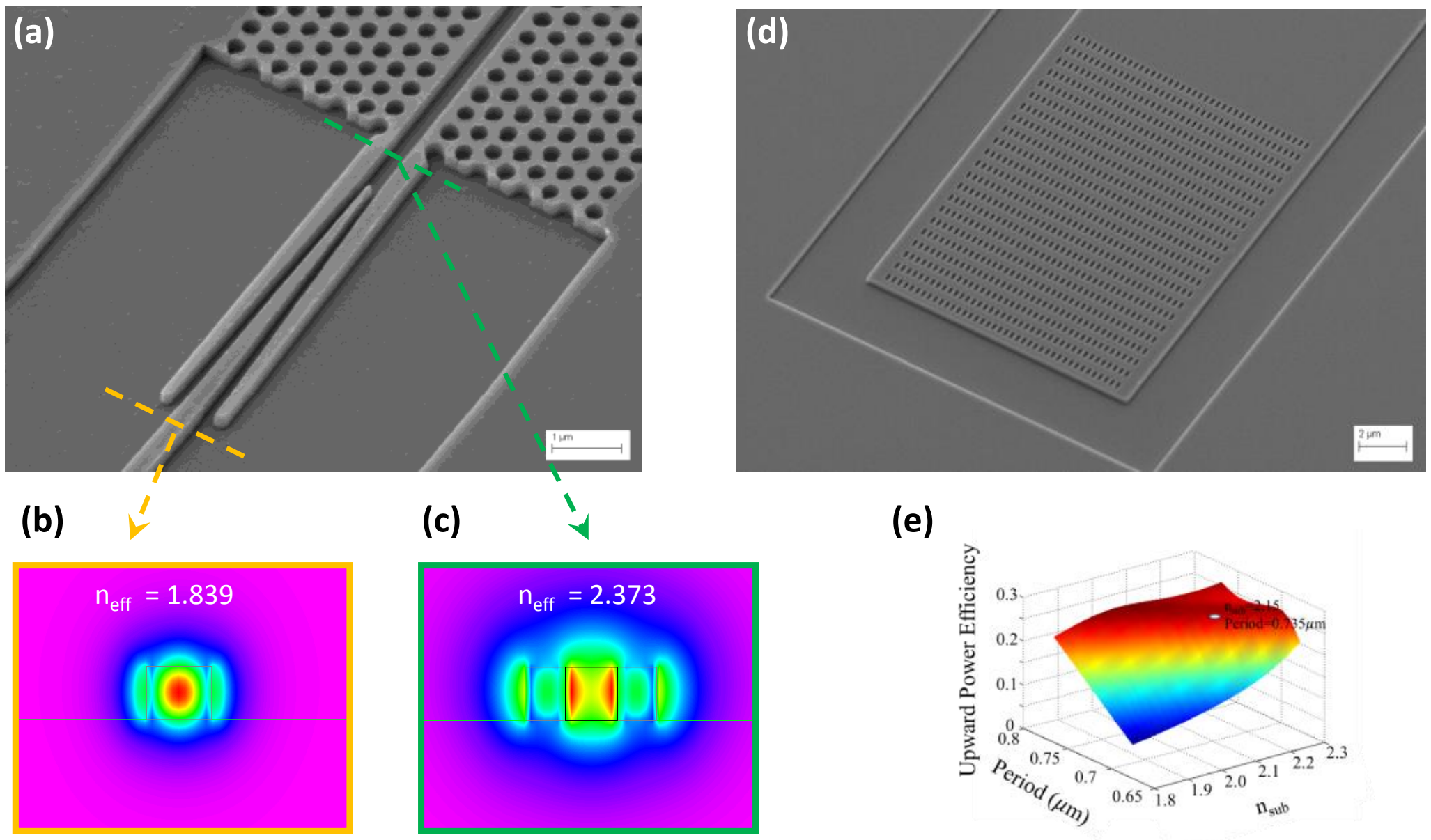

Figure 3. (a) A schematic of V-shape mode converter. (b) The optical mode profile of strip waveguide before mode converter (c) The optical mode profile of slot waveguide right after mode converter. (d) A schematic of the sub-wavelength grating coupler. (e) The coupling efficiency v.s. grating period and effective index.

## 3. RF ELECTRODE AND ANTENNA DESIGN

### 3.1 Silicon doping for high speed operation

RC time delay is one key factor limiting the broadband operation of modulators. In order for our phase modulator to operate at a high frequency, the silicon layer needs to be doped through ion implanting to reduce the resistivity while maximizing the electric field inside the slot. The doping of silicon is also beneficial for the low power consumption. The designed n-type doping levels are shown in Figure 4 (a). The resistivity values of the highly-doped silicon (donor, $10^{20}cm^{-3}$), medium-doped silicon (donor, $5\times10^{17}cm^{-3}$), and low-doped (donor, $5\times10^{16}cm^{-3}$) are $9\times10^{-6}\Omega$.m, $3\times10^{-4}\Omega$.m, and $2\times10^{-3}\Omega$.m, respectively [6]. Based on our previous experience [4], in the case of 320nm-wide slots, the resistivity of the EO polymer is about $10^{8}\Omega$.m, which is within the range of the resistivity reported for EO polymers ($\rho_{EO}$) in thin film configurations ($10^{6}\Omega$.m [7] to $10^{10}\Omega$.m [8]). The RF dielectric constant of the EO polymer ($\varepsilon_{RF,EO}$) we use here is 3.2. The change in the RF dielectric constant of silicon ($\varepsilon_{RF,Si}$) due to the doping is also taken into account [9]. Effective medium approximations are used for the calculation of both effective RF dielectric constant of silicon ($\varepsilon_{RF,eff}$) and effective resistivity of silicon ($\rho_{eff}$) in the PCW region shown in Figure 4 (a) (Hexagonal lattice, filing factor, f=0.444). For the RF dielectric constant (Electric field in X-direction), $\varepsilon_{RF,eff}$ is given as [10]

$$\varepsilon_{RF,eff} = \varepsilon_{RF,silicon} \left[ \frac{\varepsilon_{RF,EO}(1+f) + \varepsilon_{RF,Si}(1-f)}{\varepsilon_{RF,EO}(1-f) + \varepsilon_{RF,Si}(1+f)} \right] \tag{1}$$

The effective resistivity of silicon in the PCW region is estimated as

$$\rho_{eff} = \rho_{Si} \left( \frac{1+f}{1-f} \right) \tag{2}$$

where, $\rho_{si}$ is the resistivity of un-patterned silicon. Equation (2) was derived from Equations (12) and (13) in [11] assuming that the $\rho_{si} \ll \rho_{EO}$. A summary of the physical parameters used in the RF simulations is presented in Table 1.

Table 1. RF dielectric constant and conductivity values of doped silicon and doped silicon with photonic crystal patterns shown in Figure 4.

| Material | Silicon | | | | | | | | | | | | | | | | | | | | EO polymer | |
|---|---|---|---|---|---|---|---|---|---|---|---|---|---|---|---|---|---|---|---|---|---|---|
| Doping level | $1\times10^{20}$ cm$^{-3}$ | | | | $5\times10^{17}$ cm$^{-3}$ | | | | $1\times10^{17}$ cm$^{-3}$ | | | | $5\times10^{16}$ cm$^{-3}$ | | | | $1\times10^{14}$ cm$^{-3}$ | | | | | |
| Parameter | $\sigma$ | $\varepsilon_r$ | $\sigma_{eff}$ | $\varepsilon_{r,eff}$ | $\sigma$ | $\varepsilon_r$ | $\sigma_{eff}$ | $\varepsilon_{r,eff}$ | $\sigma$ | $\varepsilon_r$ | $\sigma_{eff}$ | $\varepsilon_{r,eff}$ | $\sigma$ | $\varepsilon_r$ | $\sigma_{eff}$ | $\varepsilon_{r,eff}$ | $\sigma$ | $\varepsilon_r$ | $\sigma_{eff}$ | $\varepsilon_{r,eff}$ | $\sigma$ | $\varepsilon_r$ |
| Value | $1.1\times10^5$ S/m | 22 | × | × | 2500 S/m | 12.1 | 895 S/m | 7.13 | 1111.1 S/m | 12.1 | 427.8 S/m | 7.13 | 500 S/m | 12.1 | 179 S/m | 7.13 | 2 S/m | 12.1 | 0.77 S/m | 7.13 | 1e-8 S/m | 3.2 |

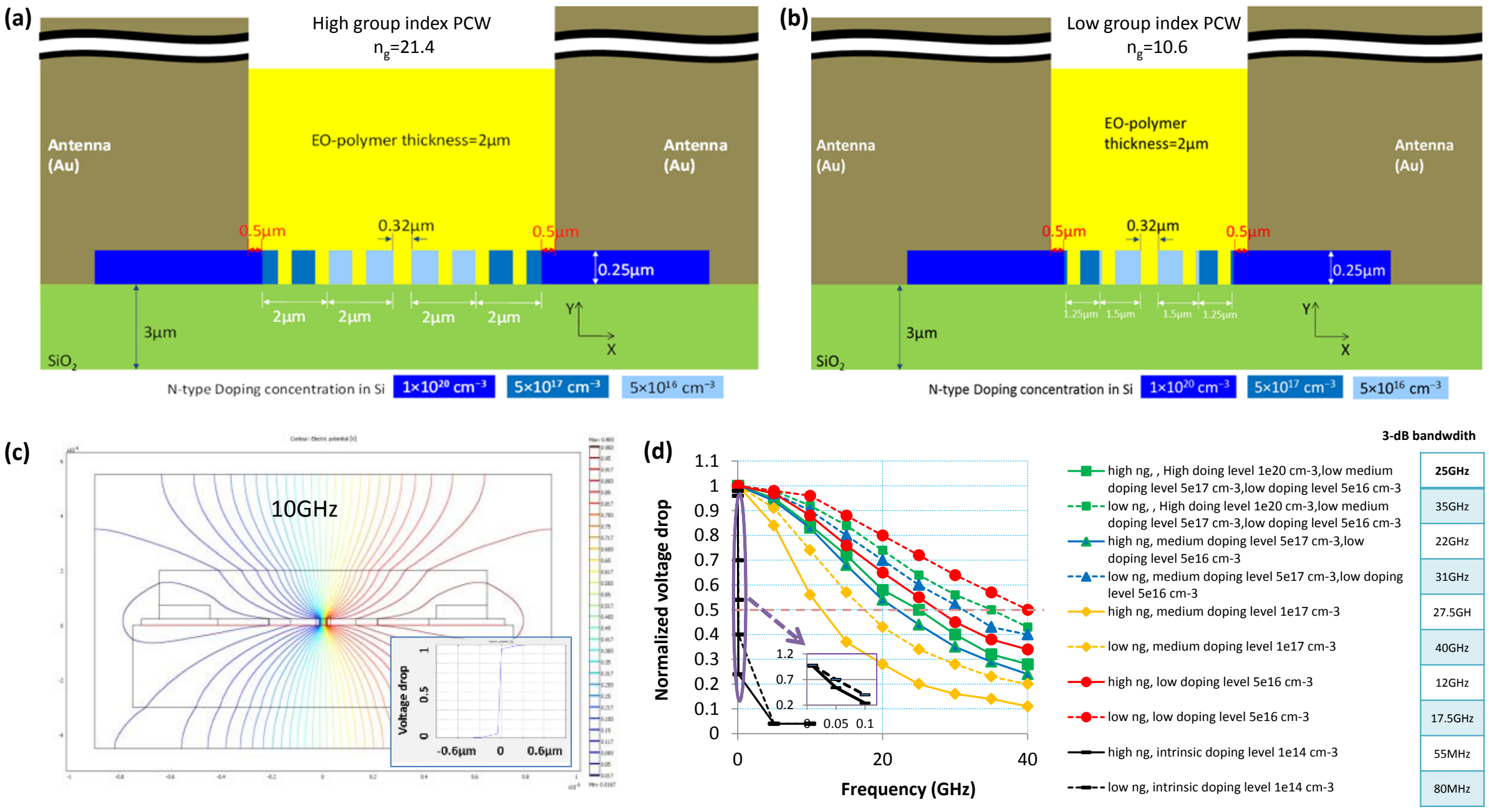

Figure 4. (a) Cross-section schematic of the antenna-coupled slot PCW ($n_g$=21.4) refilled with EO Polymer, with three levels of n-type doping concentration. (b) Cross-section schematic of the antenna-coupled slot PCW ($n_g$=10.6) refilled with EO Polymer, with three levels of n-type doping concentration. (c) RF (10GHz) Electric Potential Distribution in slot photonic crystal waveguide. This cross section view here matches the schematic shown in Figure 4 (a). Inset: electric potential along the red dashed line in (c). (d) The normalized voltage drop across the slot v.s. frequency, for different doping conditions. The 3-dB electrical bandwidths are listed.

COMSOL MULTIPHYSICS (in-Plane Electric Currents Module) is used for the RF simulations. Figure 4 (c) shows the electric potential distribution for the waveguide structure shown in Figure 4 (a), except that the antenna is replaced by perfect conductors. In Figure 4 (c), 1V potential with 10GHz frequency is applied to the electrodes. The inset of Figure 4 (c) shows that the voltage drops mostly (>90%) occurs inside the slot. As the RF frequency ($f_{RF}$) increases, the impedance of the slot ($1/(C\omega)$, where C is slot capacitance and $\omega=2\pi f_{RF}$) decreases and the fraction of the voltage dropped across the slot is reduced due to the finite resistance of silicon PCW. In order to push the operation frequency even higher, we have simulated the voltage dropped across the slot for two different devices: (1) a PCW with $n_g$=20.4 (high $n_g$, Figure 4 (a)), (2) a PCW with $n_g$=10.2 (low $n_g$, Figure 4 (b)). In the low $n_g$ PCW the lateral spread of the mode profile is smaller than that in the high $n_g$ PCW. This can reduce the distance between the electrodes (or equivalently, the antenna gap) in the case of the low $n_g$ PCW. This suggests that there is a trade-off between the RF frequency and the group index. In addition, different doping conditions are also considered. The voltage drops across the slot within a frequency range are calculated for these cases, as shown in Figure 4 (d). It can be seen that, for three-level doping condition, the 3-dB RF frequency bandwidth is 25GHz and 35GHz for the high $n_g$ and low $n_g$ PCWs, respectively. For simple doping condition, only one-level doping of $1\times10^{17}cm^{-3}$ can provide 3-dB bandwidth of 27.5GHz and 40GHz for the high $n_g$ and low $n_g$ PCWs, respectively. One disadvantage of doping is the increased optical loss. It is said in [11, 12] that silicon can be doped to about $2.5\times10^{-2}$ ohm*cm resistivity with an n-type dopant while only increasing losses approximately 5dB/cm, so the doping of $1\times10^{17}cm^{-3}$ here which gives silicon resistivity of $9\times10^{-4}$ $cm^{-3}$ is fine for low loss requirement. It is interesting to note that at low frequencies (<1MHz), the conductivity of a regular Silicon-on-Insulator (SOI) wafer (usually doped at $10^{14}cm^{-3}$) is enough for the applied voltage to entirely drop across the slot. In our previously demonstrated low frequency polymer refilled slot PCW [4] no extra doping process was performed.

### 3.2 Bowtie antenna

Both the PCW cross-section and the bowtie antenna should provide the effectiveness of the technique for operation at GHz frequency regime. The bowtie antenna designed for 10GHz operation is shown in Figure 5 (a). The designs in Figure 4 (a) together with the effective-medium approximated RF dielectric constant and conductivity values from Table 1 are used to simulate the antenna. The antenna has an interaction length of 300μm, which is the same with EO polymer refilled slot PCW length. Figure 8 (b) and (c) show the resonant field inside the antenna gap (g=8.3μm) at 10GHz. This simulation results indicate that the compressed radiation is enhanced by a factor of 15,400 inside the slot of the PCW. Figure 8 (d) shows the field enhancement (FE) factor, which is defined as the resonant electric field (inside the slot) divided by the incident field, versus the RF frequency. An unprecedented RF 1dB-bandwidth of 7GHz at 10GHz frequency can be observed. This large bandwidth has been achieved because of the small interaction length, although there is large velocity mismatch between microwaves and optical waves.

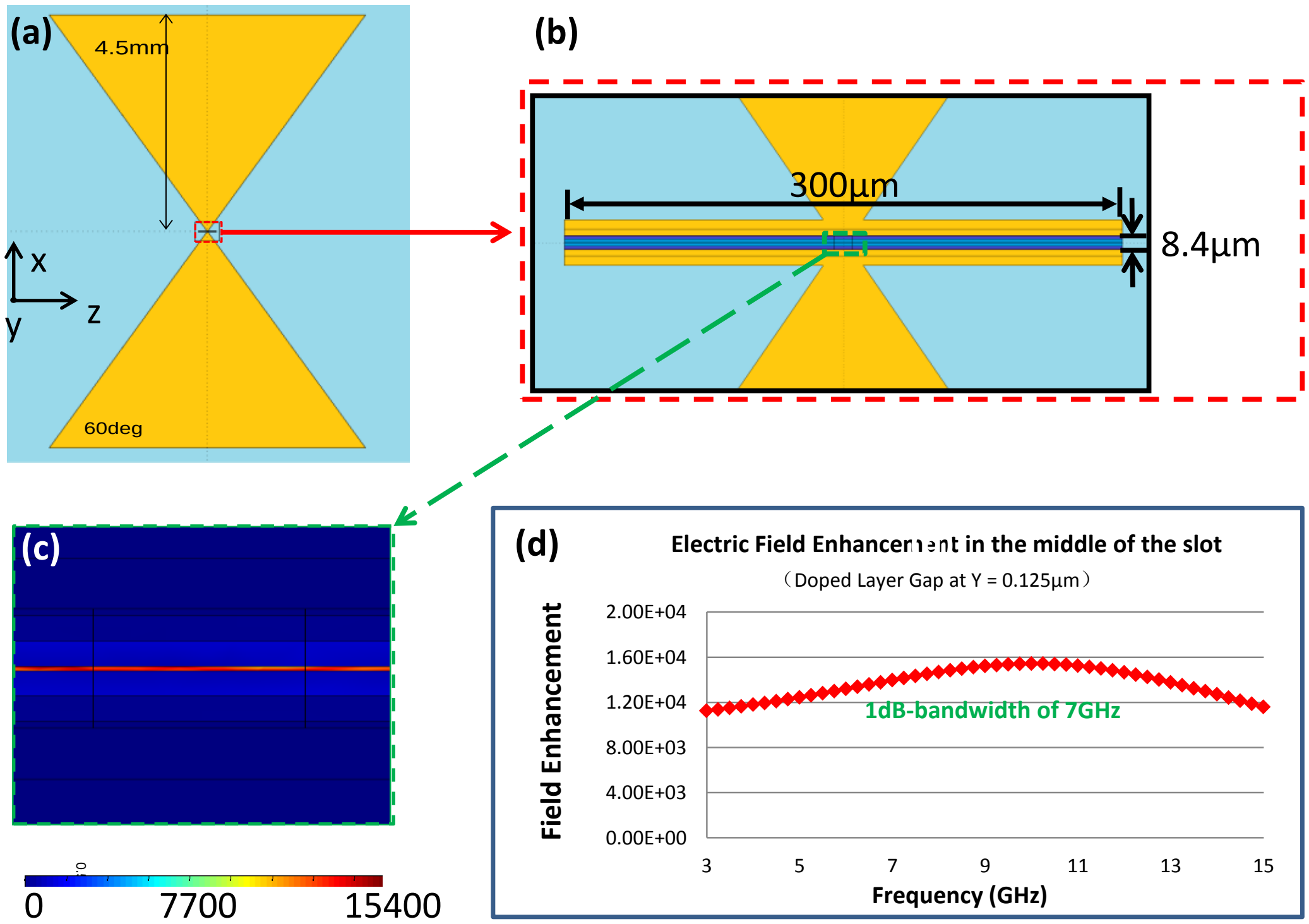

Figure 5. (a) Schematic top view of bowtie antenna. (b) Field profile inside the resonator section of the antenna. The field profile is shown inside the EO polymer refilled slot at Y=0.125μm, where Y=0 corresponds to the horizontal interface between the silicon layer and the buried oxide layer. (c) Variations of the field enhancement factor inside the slot versus the input RF frequency, indicating maximum field enhancement of 15400 at 10GHz and 1dB bandwidth of 7GHz.

## 4. CONCLUSION

We proposed an electric field sensor based on EO polymer refilled slot PCW coupled with a bowtie antenna. The slot PCW waveguide is band-engineered to achieve low-dispersion slow light propagation, maximum poling efficiency as well as optical mode confinement inside the EO polymer. The expected effective EO coefficient is 1356pm/V. Bowtie-shaped electrodes are used as both poling electrodes and as receiving antenna. The antenna is designed for operation at 10GHz and is expected to enhance the incident electric field by a factor of 15,400 inside the EO polymer refilled slot. The test results will be reported separately.

## ACKNOWLEDGEMENT

This research is supported by the AFOSR Small Business Innovation Research (SBIR) under FA8650-12-M-5131 monitored by Dr. Robert. L. Nelson.

## REFERENCES

[1] Y. A. Vlasov, and S. J. McNab, “Coupling into the slow light mode in slab-type photonic crystal waveguides,” Optics Letters, 31(1), 50-52 (2006).
[2] S. Schulz, L. O’Faolain, D. Beggs et al., “Dispersion engineered slow light in photonic crystals: a comparison,” Journal of Optics, 12(10), 104004 (2010).
[3] A. Hosseini, X. Xu, D. N. Kwong et al., “On the role of evanescent modes and group index tapering in slow light photonic crystal waveguide coupling efficiency,” Applied Physics Letters, 98(3), 031107-031107-3 (2011).
[4] X. Wang, C. Y. Lin, S. Chakravarty et al., “Effective in-device r< sub> 33</sub> of 735 pm/V on electro-optic polymer infiltrated silicon photonic crystal slot waveguides,” Optics Letters, 36(6), 882-884 (2011).
[5] C. Y. Lin, A. X. Wang, W. C. Lai et al., “Coupling loss minimization of slow light slotted photonic crystal waveguides using mode matching with continuous group index perturbation,” Optics Letters, 37(2), 232-234 (2012).
[6] S. K. Ghandhi, [VLSI fabrication principles: silicon and gallium arsenide] John Wiley & Sons, (2008).
[7] S. Huang, T. D. Kim, J. Luo et al., “Highly efficient electro-optic polymers through improved poling using a thin TiO-modified transparent electrode,” Applied Physics Letters, 96, 243311 (2010).
[8] H. Chen, B. Chen, D. Huang et al., “Broadband electro-optic polymer modulators with high electro-optic activity and low poling induced optical loss,” Applied Physics Letters, 93(4), 043507-043507-3 (2008).
[9] R. Stojan, P. Aneta, and P. Zoran, “Dependence of static dielectric constant of silicon on resistivity at room temperature,” Serbian Journal of Electrical Engineering, 1.
[10] D. Gao, and Z. Zhou, “Nonlinear equation method for band structure calculations of photonic crystal slabs,” Applied Physics Letters, 88(16), 163105-163105-3 (2006).
[11] W. Perrins, D. McKenzie, and R. McPhedran, “Transport properties of regular arrays of cylinders,” Proceedings of the Royal Society of London. A. Mathematical and Physical Sciences, 369(1737), 207-225 (1979).
[12] R. Ding, T. Baehr-Jones, Y. Liu et al., “Demonstration of a low V< sub> π</sub> L modulator with GHz bandwidth based on electro-optic polymer-clad silicon slot waveguides,” Optics Express, 18(15), 15618-15623 (2010).